\documentclass[runningheads]{llncs}
\usepackage[T1]{fontenc}
\usepackage{tabularx} 
\usepackage{graphicx}
\usepackage{adjustbox}
\usepackage{multirow}
\usepackage{makecell}
\usepackage{float}
\usepackage{caption}
\usepackage{color}
\usepackage{booktabs}
\usepackage{amsmath,amssymb,amsfonts}
\usepackage{xspace}
\usepackage{hyperref}
\usepackage[capitalise]{cleveref}

\newcommand*{\ie}{\emph{i.e.}\xspace}

\begin{document}
\title{Subcortical Masks Generation in CT Images via Ensemble-Based Cross-Domain Label Transfer}
\author{
Augustine X. W. Lee* \and
Pak-Hei Yeung* \and
Jagath C. Rajapakse
}
\institute{College of Computing and Data Science, Nanyang Technological University, Singapore \\
\email{\{alee067, pakhei.yeung\}@ntu.edu.sg}
}
\maketitle
\def\thefootnote{*}\footnotetext{These authors contributed equally to this work}\def\thefootnote{\arabic{footnote}}

\begin{abstract}
Subcortical segmentation in neuroimages plays an important role in understanding brain anatomy and facilitating computer-aided diagnosis of traumatic brain injuries and neurodegenerative disorders. However, training accurate automatic models requires large amounts of labelled data.
Despite the availability of publicly available subcortical segmentation datasets for Magnetic Resonance Imaging (MRI), a significant gap exists for Computed Tomography (CT). 
This paper proposes an automatic ensemble framework to generate high-quality subcortical segmentation labels for CT scans  
by leveraging existing MRI-based models.
We introduce a robust ensembling pipeline to integrate them and apply it to unannotated paired MRI-CT data, resulting in a comprehensive CT subcortical segmentation dataset.
Extensive experiments on multiple public datasets demonstrate the superior performance of our proposed framework.
Furthermore, using our generated CT dataset, we train segmentation models that achieve improved performance on related segmentation tasks.
To facilitate future research, we make our source code, generated dataset, and trained
models publicly available at 
\url{https://github.com/alxw0671/CT_Subcortical_Segmentation}
, marking the first open-source release for CT subcortical segmentation to the best of our knowledge.

\end{abstract}
\section{Introduction}

The human subcortex contains numerous crucial structures and regions that play crucial roles in various physiological functions underlying basic human activities \cite{janacsek2022subcortical}. For example, the thalamus facilitates the transmission of all sensory and motor signals to the cerebral cortex \cite{vertes2015thalamus} while the hippocampus is responsible for memory persistence and creation of long-term memories \cite{knierim2015hippocampus}. Besides their key physiological responsibilities, these anatomical structures have also been found to exhibit volumetric and morphological changes during the development of neurodegenerative disorders like Alzheimers’ disease \cite{yi2016relation} and Parkinson’s Disease \cite{li2022cortical}. 
Therefore, neuroimaging, which enables the analysis of the volume and morphology of the subcortical anatomies, is crucial to advance our understanding of the brain and facilitate computer-aided diagnosis of neurological conditions.

Thanks to the high contrast resolution of Magnetic Resonance Imaging (MRI) \cite{muller2002computed} that allows for better visualization of tissues, 
it has been widely adopted in subcortical analysis \cite{greve2021deep,rushmore2022anatomically}.
Specifically, there have been numerous research and methods developed for automated subcortical segmentation for MRI, with a mix of both probabilistic methods \cite{fischl2012freesurfer,puonti2016fast} and deep-learning methods \cite{henschel2020fastsurfer,billot2023synthseg,roy2019quicknat}.
In contrast, Computed Tomography (CT), another primary neuroimaging modality, has received relatively little attention in this area of research.
Despite its potential to deliver much faster (5-7 minutes \emph{vs.} 30-60 minutes for MRI) and more affordable scanning at half the cost,
the limited studies on automatic subcortical segmentation for CT have constrained its utilization in computer-aided diagnosis and treatment planning of emergent conditions, such as acute stroke or traumatic brain injuries, where CT scans are readily available.

The primary obstacle hindering the development of automated CT subcortical segmentation is the scarcity of labelled datasets.
In contrast to the abundance of publicly available labelled MRI subcortical segmentation datasets, such as the IBSR-18 \cite{ibsr18} and Mindboggle-101 \cite{klein2012101}, 
which have greatly facilitated the creation of various tools and models for this task,
there is a notable lack of similar publicly available datasets for CT subcortical segmentation.
In this work, 
we aim to fill in this gap by transferring the rich resources from the MRI community to CT, 
creating open-source and publicly available labels and pre-trained models for CT subcortical segmentation.

To achieve this goal, this paper presents a novel framework for automated CT subcortical segmentation label generation. Our framework leverages on the performance of existing MRI subcortical segmentation models and introduces a robust ensembling pipeline to integrate them.
This pipeline is then applied to a publicly available, unannotated paired MRI-CT brain dataset \cite{thummerer2023synthrad2023},
generating subcortical masks for the corresponding paired images.
Specifically, we make the following contributions:

\begin{itemize}
    \item We propose an ensemble pipeline that integrates predictions from off-the-shelf MRI subcortical segmentation tools and models. 
    Through benchmarking on multiple publicly available MRI subcortical segmentation datasets, our ensemble approach demonstrates superior performance to various state-of-the-art standalone models.
    \item We apply our proposed framework to an open-source MRI-CT brain dataset \cite{thummerer2023synthrad2023} to generate CT subcortical segmentation masks that are made publicly available. To the best of our knowledge, this constitutes the first open-access subcortical segmentation dataset for the CT modality.
    \item We train multiple segmentation models on our generated dataset and make the models and weights openly accessible.
    Extensive experiments show that the trained models exhibit accurate and robust performance in CT subcortical segmentation, as well as other tasks via transfer learning.
\end{itemize}

As the first study to make all our source codes, generated labels, and trained models publicly available for CT subcortical segmentation,
this will greatly facilitate performance benchmarking and, hence, drive development in this research area.
Although our generated subcortical segmentation labels may not be perfectly accurate due to the lack of expert manual correction, they serve as a strong prior for further refinement as future work. By releasing our trained models alongside these labels, we aim to significantly reduce the manual efforts required to annotate subcortical structures in CT images,
ultimately facilitating the development of computer-aided solutions for various CT-based downstream neuroimaging applications.

\section{Related Works}

\subsection{Whole Brain Segmentation}
Whole brain segmentation involves the partition and delineation of the brain into its respective tissue types and anatomical labels, and allows for quantitative analysis of brain tissues and structures in downstream tasks. Given MRI’s superior ability to visualize tissue contrast, whole brain segmentation algorithms are predominantly developed for MRI. Conventional probabilistic algorithms, such as FreeSurfer \cite{fischl2012freesurfer} and FIRST \cite{jenkinson2012fsl}, make use of priors from brain atlases and likelihoods from the voxel’s intensity to estimate the Maximum A Posteriori (MAP) label for each voxel, but are often limited to T1-weighted MRI. More recent probabilistic algorithms like SAMSEG \cite{puonti2016fast} also adopt the Bayesian framework to estimate MAP labels, which are able to adapt to multiple domains like both T1 and T2-weighted MRI.

While newer probabilistic algorithms have improved generalization abilities, they tend to be computationally intensive and require long processing times. The advent of deep-learning has led to the development of models for whole brain segmentation with shorter processing time. In particular, Convolutional Neural Network (CNN)-based models like FastSurfer \cite{henschel2020fastsurfer} and QuickNAT \cite{roy2019quicknat} have demonstrated commendable segmentation performance on MRI scans. Novel CNN-based methods like SynthSeg \cite{billot2023synthseg} which synthetically generates multi-contrast training data from an atlas has also shown improved generalization ability, including the capacity to segment different modalities.
Our proposed framework is designed to be applicable and agnostic to both probabilistic models and deep learning models,
ensuring its generalizability.

\subsection{CT Subcortical Segmentation}
There is much fewer research done on brain segmentation in CT modality due to the poorer tissue contrast in CT scans compared to MRI. Recent developments in CT brain segmentation include development of a 2D UNet by Cai et al. \cite{cai2020fully} which segments 11 intracranial structures and a DenseVNet by Wang et al. \cite{wang2022deep} which segments 8 brain regions. 
Despite their remarkable performance,  they primarily focus on segmenting non-subcortical structures,
with only a limited subset of subcortical structures being targeted, such as the ventricles, caudate, lentiform nucleus, internal capsule and hippocampus.
Thus, it would be meaningful to develop deep-learning models catered to subcortical segmentation.
Additionally, these studies utilized private datasets that are not publicly available, making reproducibility and performance benchmarking challenging.
In this work, we trained deep-learning models for CT subcortical segmentation, and open-sourced them and our generated dataset.

\subsection{Labelled Neuroanatomy Datasets}
The training of deep-learning models often requires large, labelled datasets. 
However, open-source neuroanatomy datasets are scarce due to patient privacy concerns and also the significant manual efforts required from expert annotators to curate these dataset. For MRI modality, some open-source segmentation datasets include the IBSR-18 \cite{ibsr18} and the MindBoggle-101 \cite{klein2012101}. In contrast, to the best of our knowledge, there is currently no open-source subcortical segmentation dataset available for CT modality. 
A study by Srikrishna et al \cite{srikrishna2021deep} shows the potential for cross-domain label propagation from MRI to CT scans. Using co-registered MRI-CT scan pairs, they carried out inference on the MRI scan using a probabilistic model and propagated the labels to the CT scan to curate a CT dataset for deep-learning training. Drawing inspiration from these prior works, we propose that open-source CT subcortical segmentation datasets can be curated using a similar approach, leveraging the extensive research conducted on MRI subcortical segmentation.

\section{Methods and Materials}

Given a dataset of $N$ pairs of unlabelled MRI-CT scans,
\(\mathcal{I} = \left\{  \mathbf{I}^{MR}_i,\mathbf{I}^{CT}_i \right\}_{i=1}^N\), where each pair consists of an $i^{th}$ MRI scan, \(\mathbf{I}^{MR}_i\), and a CT scan \(\mathbf{I}^{CT}_i\), acquired from the same patient, we propose an automated pipeline to generate subcortical segmentation labels without requiring any manual intervention.
Our framework utilizes a set of arbitrary number, $M$, of off-the-shelf MRI segmentation models, 
$\{\mathcal{S}_j(\cdot;\theta_j)\}_{j=1}^M$, 
where each model, $\mathcal{S}_j(\cdot;\theta_j)$, is parameterized by $\theta_j$.
The proposed ensembling framework, detailed in \cref{sec:method_generate}, generates robust segmentation masks, $\mathbf{L}^{MR}_i$ for the corresponding $\mathbf{I}^{MR}_i$. 
 
The generated labels are then propagated across modality from \(\mathbf{I}^{MR}\) to \(\mathbf{I}^{CT}\), as described in \cref{sec:method_propagate}.
The selection and details of the MRI segmentation models, $\mathcal{S}(\cdot;\theta)$, are outlined in \cref{sec:method_models}.
Finally, using our proposed framework, 
we generate subcortical segmentation labels, $\mathbf{L}^{CT}$ for an open-source unannotated MRI-CT paired dataset (\cref{sec:method_labels}), 
which are then used to trained different deep segmentation models (\cref{sec:method_training}).
All the generated labels and models will be made publicly available.

\subsection{Label Generation Strategy}
\label{sec:method_generate}

\begin{figure}[t]
    \centering
    \includegraphics[width=0.9\linewidth]{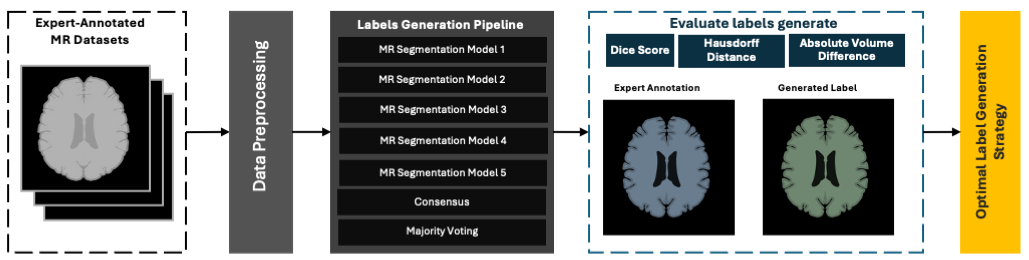}
    \caption{The workflow of our label generation strategy. It involves ensembling an arbitrary number of off-the-shelf MRI segmentation models to develop the optimal label generation strategy}
    \label{fig:label_generation_strategy}
\end{figure}

To leverage the strength of publicly available MRI subcortical segmentation tools and models, as illustrated in \cref{fig:label_generation_strategy}, we propose a label generation strategy that combines the predictions of multiple models.
For each MRI image, $\mathbf{I}^{MR} \in \mathbb{R}^{C_m \times H \times W \times D}$, 
where $C_m$, $H$, $W$ and $D$ denote the number of channels, height, width and depth, respectively,
we utilize a set of \(M\) MRI segmentation models, $\{\mathcal{S}_j(\cdot;\theta_j)\}_{j=1}^M$.
Each model outputs a subcortical predicted mask:
\begin{equation}
\label{eq:standalone}
[\mathbf{y}^{1}_i, \mathbf{y}^{2}_i, ..., \mathbf{y}^{M}_i] = [\mathcal{S}_1(\mathbf{I}^{MR}_i;\theta_1), \mathcal{S}_2(\mathbf{I}^{MR}_i;\theta_2), ..., \mathcal{S}_M(\mathbf{I}^{MR}_i;\theta_M)],
\end{equation}
where each prediction, $\mathbf{y} \in \mathbb{R}^{C_c \times H \times W \times D}$ has $C_c$ classes, height $H$, width $W$ and depth $D$.
We then integrate these predicted masks into the final labels using our proposed ensembling approaches.
Noted that many off-the-shelf MRI subcortical segmentation models only provide hard segmentation outputs (\ie 0 and 1) without access to the inner layers of the model (\ie soft probability scores).
To ensure the model-agnostic nature and generalizability of our proposed framework, 
we perform ensembling on the hard segmentation outputs of the models. We explore two ensembling methods: consensus and majority voting in this work.

Both approaches require getting a count map, $\mathbf{C}_i \in \mathbb{R}^{C_c \times H \times W \times D}$, for its corresponding $\mathbf{I}^{MR}_i$ by:
\begin{equation}
\label{eq:count}
\mathbf{C} = \sum_{j=1}^{M} \mathbf{y}_i^j.
\end{equation}
For every voxel, 
$\mathbf{v} = \mathbf{C}(:,x,y,z)$,
where $\mathbf{v} \in \mathbb{R}^{C_c}$,
we identify the class with the most predictions, $c_{max}$, as:
\begin{equation}
\label{eq:argmax}
c_{max} = \arg\max_{c \in C_c} \mathbf{v}(c).
\end{equation}
The consensus ensembling, $f_{concensus}(\cdot)$, classifies each voxel $\mathbf{v}$ as a particular class only if all models, $\{\mathcal{S}_j(\cdot;\theta_j)\}_{j=1}^M$ agree on that class; otherwise it is classified as the background class, $bg$. This is formulated as:
\begin{equation}
\label{eq:consensus}
f_{\text{consensus}}(\mathbf{v}) = 
\begin{cases}
c_{max} & \text{if } \mathbf{v}(c_{max}) = M, \\
bg & \text{else}.
\end{cases}
\end{equation}
However, this approach can be overly strict, resulting in a strong bias towards the background class. 
To address this, we propose majority voting as an improvement by imposing a less strict rule. 
This method classifies each voxel $\mathbf{v}$ as the class that receives the most votes, namely $c_{max}$. In cases of ties where multiple classes have the same maximum count, 
the voxel is classified as the background class.
The final subcortical mask, $\mathbf{L}^{MR}_i \in \mathbb{R}^{H \times W \times D}$, corresponding to $\mathbf{I}^{MR}_i$ is generated by combining the prediction of every voxel $\mathbf{v}$.

\subsection{Label Propagation from MRI to CT}
\label{sec:method_propagate}

\begin{figure}[t]
    \centering
    \includegraphics[width=0.9\linewidth]{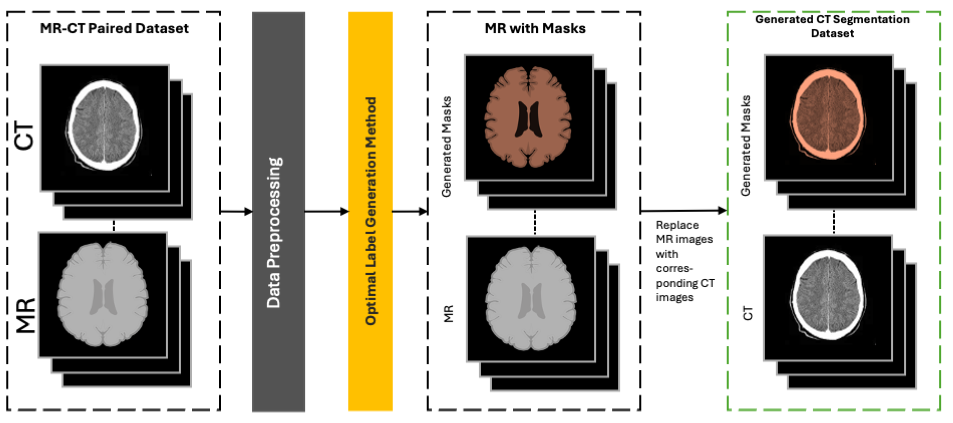}
    \caption{The workflow of our label propagation method. For an MRI-CT scan pair, the optimal label generation strategy is applied on an MRI scan, $\mathbf{I}^{MR}_i$ and subsequently propagated to the co-registered CT scan, $\mathbf{I}^{CT}_i$}
    \label{fig:label_propagation}
\end{figure}

Given that each MRI-CT scan pair, $\mathbf{I}^{MR}_i$ and $\mathbf{I}^{CT}_i$, is from the same patient, we can infer that they share identical regions of interest.
Therefore, the generated segmentation label, $\mathbf{L}^{MR}_i$, obtained in \cref{sec:method_generate} for the MRI scan, $\mathbf{I}^{MR}_i$, can be transferred to the corresponding CT scan, $\mathbf{I}^{CT}_i$, as illustrated in \cref{fig:label_propagation}. 

To achieve this, we first compute the registration between the paired images by finding the optimal spatial transformation, $\hat{T}_i$, to align $\mathbf{I}^{MR}_i$  with $\mathbf{I}^{CT}_i$:
\begin{equation}
\label{eq:registration}
\hat{T}_i = \arg\min_{T} \; \mathcal{C}_{\text{sim}}\left(\mathbf{I}^{CT}_i, \mathbf{I}^{MR}_i \circ T\right), 
\end{equation}
where $\mathcal{C}_{\text{sim}}$ is the similarity cost function, such as mean square error.
\cref{eq:registration} is a simplified generalization of the registration process.
In practice, regularization terms that encourage smooth or diffeomorphic transformations may be added.
The details of the image registration are beyond the scope of this work, and our proposed framework is agnostic to the choice of registration methods.
Off-the-shelf image registration tools \cite{klein2009elastix} or deep learning-based registration approaches \cite{cao2018deep} can be employed for this purpose.

Once the optimal spatial transformation, $\hat{T}_i$, is computed, the subcortical segmentation label, $\mathbf{L}^{CT}_i$, of the corresponding CT scan, $\mathbf{I}^{CT}_i$, can be generated by transforming $\mathbf{L}^{MR}_i$ with $\hat{T}_i$:
\begin{equation}
\label{eq:propagation}
\mathbf{L}^{CT}_i = \mathbf{L}^{MR}_i \circ \hat{T}_i. 
\end{equation}

\subsection{Choices of MRI Segmentation Models}
\label{sec:method_models}
Our proposed framework can accommodate an 
arbitrary number, $M$, of off-the-shelf MRI segmentation models, 
$\{\mathcal{S}_j(\cdot;\theta_j)\}_{j=1}^M$, as explained in \cref{sec:method_generate}.
In this study,
we implemented five publicly available MRI segmentation models, $\mathcal{S}(\cdot;\theta)$. 
The selected models include both probabilistic-based and deep-learning-based approaches, 
with their details as follows:
\\
\\
\noindent \textbf{ASeg} \cite{fischl2012freesurfer}
is the default subcortical segmentation model employed by Freesurfer recon-all pipeline \cite{fischi2002whole}. It is a probabilistic atlas-based segmentation model.
\\
\\
\noindent \textbf{Sequence Adaptive Multimodal Segmentation (SAMSEG)} \cite{puonti2016fast}
is a probabilistic model that utilizes Bayesian modelling for whole brain segmentation. It is also packaged within the Freesurfer toolkit.
\\
\\
\noindent \textbf{FastSurfer} \cite{henschel2020fastsurfer}
is a deep-learning-based whole-brain segmentation model built on a 2D UNet. It serves as an alternative to Freesurfer and reduces the processing time significantly.
\\
\\
\noindent \textbf{SynthSeg} \cite{billot2023synthseg}
is a CNN-based model designed to perform segmentation on brain scans across various domains and contrasts through synthetic generation of a wide range of training data.
\\
\\
\noindent \textbf{QuickNAT} \cite{roy2019quicknat}
is a CNN-based model optimized for fast brain segmentation. It is trained on labels from multiple segmentation softwares and fine-tuned with manually-annotated data.

\subsection{Generation of CT Subcortical Segmentation Labels}
\label{sec:method_labels}
Using our proposed framework, as summarized in \cref{sec:method_generate,sec:method_propagate,sec:method_models}, we generated CT subcortical segmentation labels, $\mathbf{L}^{CT}$, for an unannotated MRI-CT paired dataset.
The dataset used was the open-access paired MRI-CT brain dataset \cite{thummerer2023synthrad2023} released by the \textbf{SynthRAD Grand Challenge 2023} \cite{huijben2024generating}. 
This dataset consists of scans from 180 subjects, evenly distributed across three different medical centers. 

The dataset provides both T1-weighted MRI and CT scans for each subject, with the MRI and CT scans already aligned with each other. 
The scans were preprocessed by cropping to the bounding box defined by the patient's outline, with a 20-voxel margin.
We selected 17 subcortical regions to include in our segmentation dataset, based on their physiological significance: Lateral Ventricles (L/R), Thalamus (L/R), Caudate (L/R), Putamen (L/R), Pallidum (L/R), Hippocampus (L/R), Amygdala (L/R), Accumbens Area (L/R) and Brainstem.
Although other subcortical regions, such as the Substantia Nigra, have key physiological roles, they are too small and not all MRI segmentation models provide segmentation masks for them.

Both MRI and CT scans were resampled to a resolution of $1 \times 1 \times 1 mm^3$, with a maximum image dimension of $256 \times 256 \times 256$.
The MRI scans underwent additional processing using Freesurfer's autorecon1 pipeline, which includes intensity correction, Talairach transformation to the MNI305 atlas and intensity normalization.

Since the paired MRI-CT images in this dataset were already aligned using rigid image registration with Elastix \cite{klein2009elastix},
we could skip the registration steps described in \cref{eq:registration,eq:propagation}. 
Instead, we directly transferred the generated $\mathbf{L}^{MR}$ to obtain $\mathbf{L}^{CT}$.
The resulting set of CT subcortical labels is publicly available at \url{https://github.com/alxw0671/CT_Subcortical_Segmentation}.

\subsection{Training of CT Subcortical Segmentation Models}
\label{sec:method_training}
Using the CT subcortical labels generated in \cref{sec:method_labels} and their corresponding CT images,
we trained numerous CNN-based and Transformer-based segmentation models. The details of these models are as follows:
\\
\\
\noindent \textbf{UNet}\cite{ronneberger2015u}
is a CNN-based model characterized by a series of encoders and decoders connected by skip connections, forming a U-shaped architecture. 
The UNet has achieved impressive performance in various medical segmentation tasks \cite{azad2024medical}. In this study, we trained both 2D and 3D versions of UNet.
\\
\\
\noindent \textbf{SwinUNETR}\cite{hatamizadeh2021swin}
differs from conventional UNets by using Swin Transformers \cite{liu2021swin} as its encoders instead of convolutional layers. This design enables the model to capture long-range global context more effectively.
\\
\\
\noindent \textbf{nnUNet}\cite{isensee2021nnu}
is a state-of-the-art model that features a self-configuring pipeline that automatically trains a UNet with an optimal parameter configuration, eliminating the need for manual hyperparameter tuning.

\section{Experimental Setup}

\subsection{Optimal Label Generation Strategy}
\label{sec:exp_generate}

To evaluate the performance of our proposed ensemble framework, we compared it with each of the off-the-shelf MRI segmentation models introduced in \cref{sec:method_models} using two publicly available expert-annotated datasets  (detailed in \cref{sec:exp_datasets}).
We implemented the two ensembling methods, consensus and majority voting, as described in \cref{sec:method_generate} for our proposed framework and benchmarked their performance.

\subsection{CT Subcortical Segmentation Models}
\label{sec:exp_models}

We trained both CNN-based and Transformer-based models, as introduced in \cref{sec:method_training}, on our generated CT subcortical segmentation dataset.  
The dataset was split into training (70\%), validation (15\%), and test sets (15\%). 
To enhance model performance, 
we applied additional preprocessing steps, combining the left and right regions of each structure.

We trained the SwinUNETR \cite{hatamizadeh2021swin}, imported from the MONAI library \cite{cardoso2022monai}, and the UNets using the PyTorch framework.
They were optimized using Dice loss and the Adam \cite{kingma2014adam} optimizer with an initial learning rate of 0.0001.
The learning rate was decayed using the ReduceLROnPlateau scheduler with a patience of 3, based on the validation loss.
Early stopping was implemented with a patience of 5.
The nnUNet \cite{isensee2021nnu} was trained using the \textit{nnUNet v2} framework \cite{isensee2021nnu} with the \textit{3d\_fullres} configuration from the official source codes\footnote{https://github.com/MIC-DKFZ/nnUNet}. The training process consisted of 1000 epochs and hyperparameters tuning was automatically performed by the framework.
\subsection{Transfer Learning}
The scarcity of publicly available CT subcortical segmentation datasets poses a challenge in directly evaluating the quality of our generated segmentation dataset.
To address this limitation, we proposed transfer learning as an indirect yet practical method of validating our segmentation labels.

Using a 3D UNet trained on our generated CT subcortical segmentation dataset, as described in \cref{sec:method_training,sec:exp_models},
we froze its encoder and fine-tuned its decoder on an open-source, expert-annotated MRI dataset, OASIS-TRT-20 \cite{klein2012101}, under limited data conditions by only using 5 annotated scans for training. The details of the OASIS-TRT-20 dataset are provided in \cref{sec:exp_datasets}.

To assess the effectiveness of the features learned from our generated dataset, we compared the fine-tuned model with a 3D UNet trained from scratch on the same MRI dataset. This comparison allowed us to evaluate the transferability of the knowledge learned from our CT subcortical segmentation dataset to a different modality (MRI) and dataset. To ensure the results were not due to model bias, we also conducted the transfer learning experiment with ResUNet.

\subsection{Evaluation Datasets}
\label{sec:exp_datasets}
We utilized two MRI subcortical segmentation datasets in our experiments. 
Both datasets' voxel spacing was preprocessed to have a uniform voxel spacing of $1mm^3$ and cropped to a maximum dimension of $256\times256\times256mm^3$ to standardize the outputs of various MRI segmentation models.
\\
\\
\textbf{The IBSR-18 dataset} \cite{ibsr18} contains 18 manually-guided annotated T1-weighted MRI brain scans from 18 healthy subjects. 
The dataset was provided by the Center for Morphometric Analysis at Massachusetts General Hospital\footnote{http://www.cma.mgh.harvard.edu/ibsr/}. 
The original scans and masks have dimensions of $256\times256\times128$ with the voxel spacing of $0.9375\times0.9375\times1.5 mm^3$.
\\
\\
\noindent \textbf{The OASIS-TRT-20 dataset} \cite{klein2012101} is part of the Mindboggle-101 project and contains 20 T1-weighted MRI brain scans from 20 healthy subjects aged between 23-29 years old.
 
\subsection{Evaluation Metrics}
\label{sec:exp_metrics}
The segmentation performance of different methods in our experiments was evaluated by 3 metrics: Dice-Sørensen coefficient (DSC), undirected Hausdorff Distance (HD) and Absolute Volume Difference (AVD).

\section{Results}
\subsection{Optimal Label Generation Strategy}

\begin{table}[t]
\caption{Segmentation results on two MRI expert-annotated datasets}
\label{tab:label_generation_table}
\centering
\footnotesize
\begin{tabular}{l|l|l|l|l|l|l}
\hline
\multirow{2}{*}{\textbf{Label Generation Method}} & \multicolumn{3}{c|}{\textbf{IBSR-18}} & \multicolumn{3}{c}{\textbf{OASIS-TRT-20}} \\
\cline{2-7}
 & \makecell{\textbf{DSC} \\ \ensuremath{\uparrow}}
 & \makecell{\textbf{HD} \\ (vox)\ensuremath{\downarrow}}
 & \makecell{\textbf{AVD} \\ (vox)\ensuremath{\downarrow}}
 & \makecell{\textbf{DSC} \\ \ensuremath{\uparrow}}
 & \makecell{\textbf{HD} \\ (vox)\ensuremath{\downarrow}}
 & \makecell{\textbf{AVD} \\ (vox)\ensuremath{\downarrow}}\\
\hline
ASeg (FreeSurfer) \cite{fischl2012freesurfer}                & 0.796 & 4.8  & 415.2  & 0.785 & 6.2  & 602.3  \\
SAMSEG \cite{puonti2016fast}                           & 0.796 & 4.9  & 658.8  & 0.758 & 6.7  & 1312.3 \\
FastSurfer \cite{henschel2020fastsurfer}                       & 0.820 & 4.6  & 397.5  & 0.802 & 6.0  & 697.0  \\
SynthSeg \cite{billot2023synthseg}                         & 0.824 & 4.4  & 357.1  & 0.806 & 5.2  & 883.2  \\
QuickNAT \cite{roy2019quicknat}                         & 0.834 & 10.7 & 455.9  & 0.795 & 32.4 & 812.0  \\
\hline
Consensus (All Models)          & 0.784 & 5.6  & 995.3  & 0.772 & 6.6  & 1579.1 \\
Consensus (Deep Learning Models) & 0.826 & 5.1  & 648.5  & 0.807 & 6.1  & 1040.0 \\
Majority Voting (All Models)    & 0.845 & 4.1  & 312.5  & 0.820 & 5.0  & 544.9  \\
Majority Voting (Deep Learning Models) & \textbf{0.852} & \textbf{4.0}  & \textbf{312.5}  & \textbf{0.825} & \textbf{5.0}  & \textbf{500.4}  \\
\hline
\end{tabular}%
\vspace{-10pt}
\caption*{\hspace{0pt}\parbox[t]{0.9\linewidth}
\footnotesize{
\textit{$\uparrow$ means higher values being more accurate} \\
\textit{\textbf{Bold} indicates the best performance}
}
}
\end{table}

We compared the performance of our proposed framework with each of the off-the-shelf MRI segmentation models introduced in \cref{sec:method_models}.
The overall results are presented in \cref{tab:label_generation_table}, 
while detailed results for each subcortical structure on both datasets can be found in the Supplementary Tables 1-6.

As shown in \cref{tab:label_generation_table}, when used as standalone models, deep learning-based approaches
like QuickNAT and SynthSeg
tended to generate labels with higher overlap with the ground-truth, as evidenced by their higher average DSC for both datasets. This can be attributed to the ability of deep models to learn complex representations.  
However, they did not necessarily exhibit greater robustness than probabilistic models, as demonstrated by QuickNAT's significantly higher HD for both datasets.
Notably, no single model consistently achieved the highest DSC and lowest HD and AVD.

In contrast, 
our proposed framework, which employs majority voting ensembling, demonstrated superior and more robust and consistent performance.
This was reflected in its higher DSC and lower HD and AVD compared to all other methods.
While the strict rule of consensus ensembling may result in smaller integrated segmentation labels,
leading to poorer results,
majority voting ensembling improves on this by eliminating outliers specific to a minority of the models without significantly shrinking the segmented volume.

To further improve robustness, we evaluated the performance of majority voting ensembling using only deep learning models. As expected, given their higher DSC values, this approach generated labels with the highest average DSC and lowest HD and AVD for both datasets. Our results demonstrated that our proposed framework, which leverages majority voting ensembling, produces more robust segmentation masks than any individual model, proving the effectiveness of our proposed framework.

\subsection{CT Subcortical Segmentation Models}

\begin{table}[t]
\caption{Average DSC of segmentation of different structures by various deep models trained on our generated CT subcortical segmentation dataset}
\label{tab:CT_segmentation_results}
\resizebox{\textwidth}{!}{%
\begin{tabular}{@{}l|lllllllll@{}}
\hline
Model & Ventricles & Thalamus & Caudate & Putamen & Pallidum & Hippocampus & Brainstem & Average \\ \hline
2D UNet   & 0.878      & 0.896    & 0.834   & 0.814   & 0.789    & 0.754       & 0.907     & 0.839   \\
SwinUNETR & 0.884      & 0.902    & 0.853   & 0.843   & 0.830    & 0.771       & 0.911     & 0.856   \\
3D UNet   & 0.893      & 0.910    & 0.856   & 0.844   & 0.837    & 0.777       & 0.917     & 0.862   \\
\textbf{nnUNet} & \textbf{0.914} & \textbf{0.935} & \textbf{0.896} & \textbf{0.895} & \textbf{0.883} & \textbf{0.864} & \textbf{0.949} & \textbf{0.905}   \\ \hline
\end{tabular}%
}
\end{table}

\begin{figure}[]
    \centering
    \includegraphics[width=0.9\linewidth]{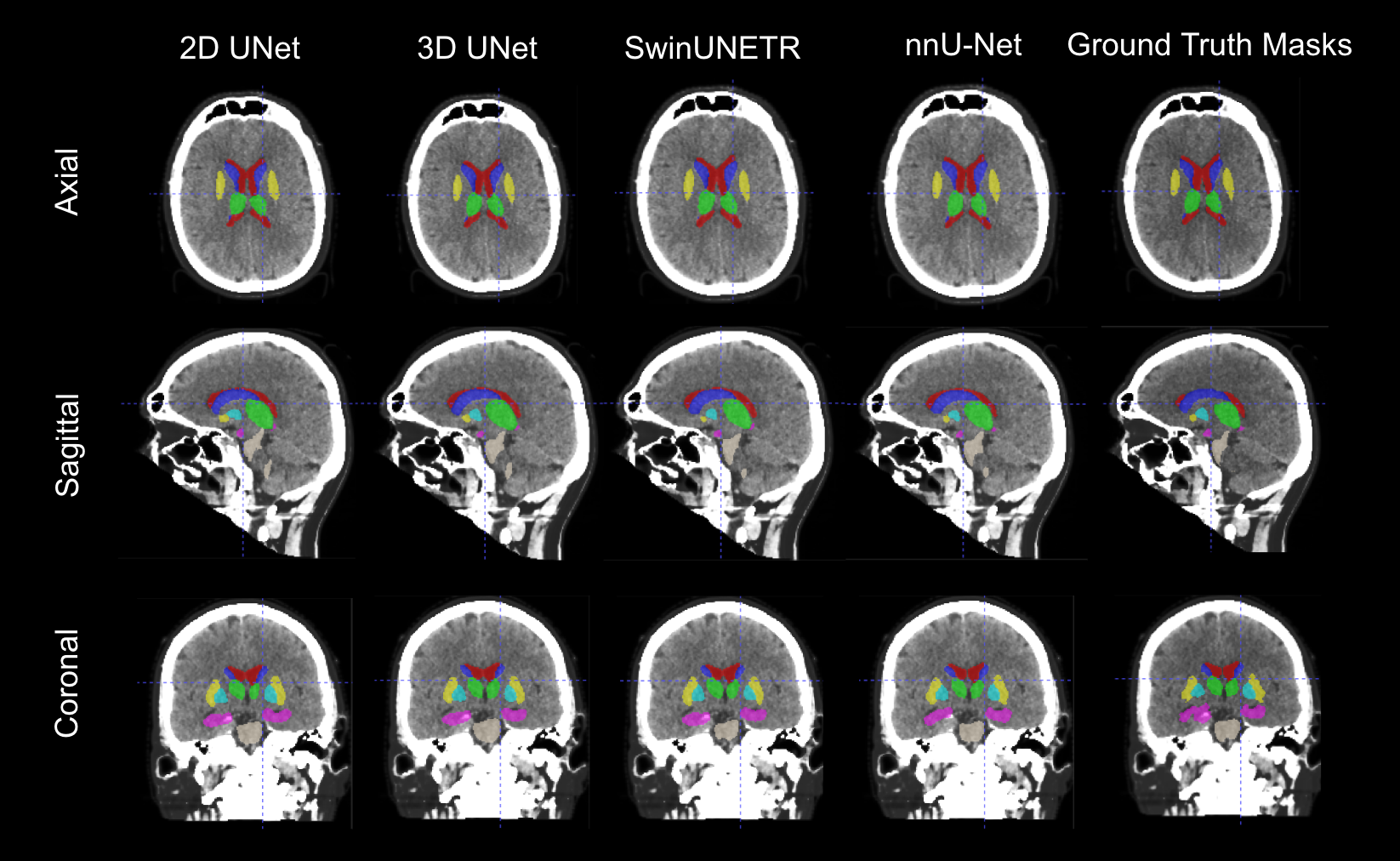}
    \caption{Qualitative results by various CT subcortical segmentation models. The results and ground-truth across the axial, sagittal and coronal axes.}
    \label{fig:enter-label}
\end{figure}

We further evaluated the performance of different models trained on our generated CT subcortical segmentation dataset. The qualitative results are shown in \cref{fig:enter-label}, while the quantitative results are presented in \cref{tab:CT_segmentation_results}.

As shown in \cref{tab:CT_segmentation_results}, CNN-based models, namely 3D UNet and nnUNet, outperformed Transformer-based model, SwinUNETR. 
This is likely attributed to the limited amount of training data, which may not be sufficient to fully leverage the capabilities of the transformer-based architecture.
Nevertheless, our trained models have established a performance baseline for future works aiming to improve the performance of segmentation models for CT subcortical segmentation.

\subsection{Validating dataset's utility through Transfer Learning}

\begin{figure}[t]
    \centering
    \includegraphics[width=0.9\linewidth]{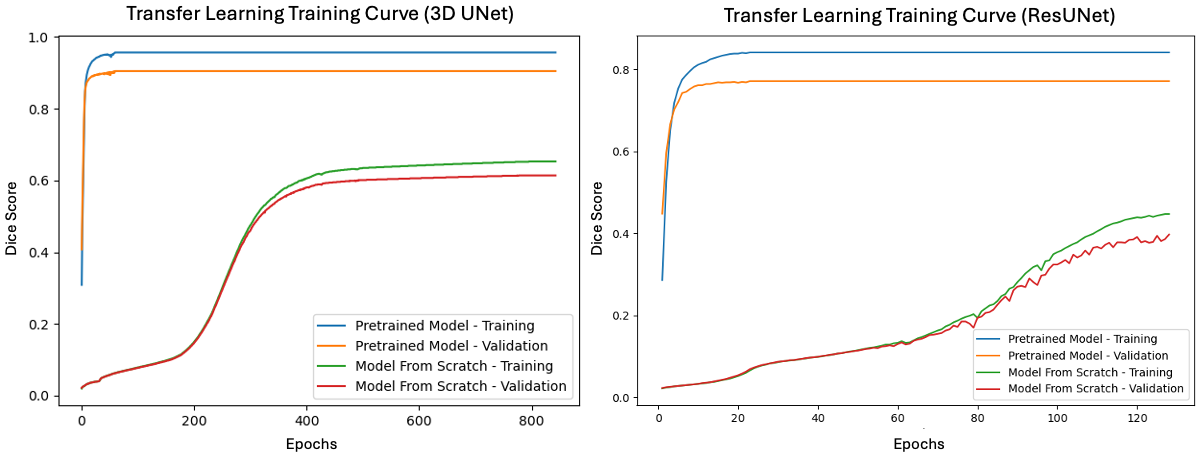}
    \caption{Training and Validation Dice Scores of both pretrained models and models trained from scratch.}
    \label{fig:training_curve}
\end{figure}

\begin{table}[t]
\caption{Evaluation of segmentation performance of both pretrained model and model trained from scratch using transfer learning}
\label{tab:transfer_learning_results}
\centering
\resizebox{\textwidth}{!}{
\begin{tabular}{l|l|l|l|l|l|l|l|l|l|l|l|l}
\hline
\multirow{2}{*}{Subcortical Structure} & \multicolumn{3}{c|}{Pretrained UNet} & \multicolumn{3}{c|}{UNet from Scratch} & \multicolumn{3}{c|}{Pretrained ResUNet} & \multicolumn{3}{c}{ResUNet from Scratch} \\
\cline{2-13}
 & \makecell{DSC \\ \ensuremath{\uparrow}}
 & \makecell{HD \\ (vox)\ensuremath{\downarrow}}
 & \makecell{AVD \\ (vox)\ensuremath{\downarrow}}
 & \makecell{DSC \\ \ensuremath{\uparrow}}
 & \makecell{HD \\ (vox)\ensuremath{\downarrow}}
 & \makecell{AVD \\ (vox)\ensuremath{\downarrow}}
 & \makecell{DSC \\ \ensuremath{\uparrow}}
 & \makecell{HD \\ (vox)\ensuremath{\downarrow}}
 & \makecell{AVD \\ (vox)\ensuremath{\downarrow}}
 & \makecell{DSC \\ \ensuremath{\uparrow}}
 & \makecell{HD \\ (vox)\ensuremath{\downarrow}}
 & \makecell{AVD \\ (vox)\ensuremath{\downarrow}} \\
\hline
Ventricles & 0.905 & 21.3  & 1921.2  & 0.936 & 18.2  & 636.1 & 0.853 & 55.5  & 3922.0  & 0.810 & 48.8  & 4796.1  \\
Thalamus & 0.931 & 5.3  & 551.9  & 0.943 & 2.6  & 416.3 & 0.906 & 21.8  & 929.9  & 0.021 & 75.7  & 1908400.0 \\
Caudate & 0.896 & 15.8  & 504.3  & 0.935 & 19.3  & 146.5 & 0.856 & 24.8  & 621.9  & 0.798 & 45.2  & 1174.2  \\
Putamen & 0.912 & 37.1  & 541.5  & 0.939 & 6.8  & 155.9 & 0.861 & 39.0  & 704.9  & 0.814 & 45.4  & 2516.0  \\
Pallidum & 0.896 & 3.9 & 370.2  & 0.622 & 24.8 & 5388.1 & 0.849 & 3.1 & 452.9  & 0.000 & 30.9 & 4882.0  \\
Hippocampus & 0.868 & 14.3  & 533.3  & 0.882 & 6.2  & 440.6 & 0.774 & 38.6  & 1163.9  & 0.000 & 46.1  & 10100.1 \\
Brainstem & 0.947 & 11.6  & 392.8  & 0.957 & 4.4  & 317.8 & 0.899 & 27.9  & 2506.8  & 0.754 & 38.3  & 8514.2 \\
Amygdala & 0.857 & 8.9  & 140.5  & 0.001 & 230.9  & 6920161.5 & 0.004 & 102.6  & 1968550.2  & 0.000 & 98.0  & 3576.4 \\
Accumbens Area & 0.839 & 23.7  & 168.7  & 0.000 & 5.6  & 1477.7 & 0.736 & 17.0  & 154.4  & 0.000 & 10.4  & 1477.8 \\ \hline
Average & \textbf{0.895} & \textbf{15.8}  & \textbf{569.4}  & 0.691 & 35.4 & 769904.5 & \textbf{0.749} & \textbf{36.7}  & 219889.7  & 0.355 & 48.8 & \textbf{216159.6} \\ \hline
\end{tabular}
}
\\
\footnotesize{\textit{$\uparrow$ means higher values indicate better segmentation performance}}
\end{table}

Finally, we assessed the utility of our generated CT subcortical segmentation dataset by pretraining a 3D UNet and a ResUNet on our CT dataset and fine-tuning them with a small amount (\ie 5) of annotated MRI images, followed by comparing them to the same networks which were trained from scratch.

As illustrated in \cref{fig:training_curve}, the training curves of 3D UNet reveal that the pretrained model converged significantly faster at 65 epochs, whereas the model trained from scratch required more than 840 epochs to converge. Similarly, the pretrained ResUNet converged much faster at 28 epochs while the ResUNet trained from scratch required more than 130 epochs. Additionally, the validation Dice score for the pretrained models is significantly higher, suggesting its better performance. To further evaluate their segmentation capabilities, we applied the models to the test dataset, and the results are presented in \cref{tab:transfer_learning_results}.
Notably, for the three smallest structures, Pallidum, Amygdala and Accumbens Area, the pretrained model performed significantly better than the model trained from scratch, leading to higher overall segmentation accuracy.

The faster convergence speed and superior segmentation performance of the pretrained model indirectly validate the quality and utility of our generated CT subcortical segmentation labels, suggesting that it can serve as a strong reference standard for training deep-learning models. 
Our transfer learning experiments further demonstrate the potential of our dataset to facilitate the training of deep models for related medical image analysis tasks with limited annotated data. This is particularly useful in practice, where acquiring expert-annotated data can be resource-intensive and challenging.

\section{Conclusion}
In summary, we have proposed an automated ensemble framework that leverages existing MRI segmentation models to generate robust and accurate segmentation labels for CT scans. This framework effectively addresses the data scarcity problem in CT subcortical segmentation and greatly reduces the manual annotation effort required by clinical experts. As a model-agnostic pipeline, it can be easily extended to incorporate future improvements in segmentation, further enhancing its robustness. 
By utilizing this pipeline, we have generated an open-source CT subcortical segmentation dataset and trained reliable segmentation models on it, providing a strong foundation for future research and performance benchmarking. 
Potential avenues for future work include extending the framework to generate labels for additional subcortical anatomies beyond the 17 classes currently addressed, as well as exploring its applicability to other imaging modalities. Semi-automated and community-driven label correction methods can also be explored and incorporated to further enhance the labels' robustness.

\subsubsection{Acknowledgments.} PH. Yeung is funded by the Presidential Postdoctoral Fellowship from Nanyang Technological University.

\subsubsection{Disclosure of Interest.} The authors have no competing interests to declare.



\bibliographystyle{splncs04}
\bibliography{ref}

\end{document}